\documentclass[twocolumn,aps,prl,superscriptaddress]{revtex4-2}
\usepackage{amsfonts}
\usepackage{amsmath}
\usepackage{amssymb}
\usepackage{graphicx}
\usepackage{braket}

\usepackage[colorlinks, urlcolor=cyan, citecolor=blue, linkcolor=magenta]{hyperref}

\begin{document}

\title{Experiment-compatible measurement--feedback quantum state preparation with reinforcement learning}
\author{Xiaotian Nie}
\affiliation{Intelligent Quantum Inception Co., Ltd., Haidian, Beijing, 100083, China}
\author{Tao Zhang}
\email{tao\_zhang@mail.tsinghua.edu.cn}
\affiliation{Institute for Advanced Study, Tsinghua University, Beijing 100084, China}
\affiliation{Intelligent Quantum Inception Co., Ltd., Haidian, Beijing, 100083, China}
\author{Linghui Chen}
\email{lhchen@iflytek.com}
\affiliation{iFLYTEK Research, Hefei, 230088, China}
\affiliation{Intelligent Quantum Inception Co., Ltd., Haidian, Beijing, 100083, China}

\date{\today}

\begin{abstract}
Ground-state preparation is a critical task in quantum simulation and quantum computing, as it enables the study of correlated phases and the generation of entangled resource states. While measurement--feedback control has emerged as a promising route to state preparation, existing schemes either rely on handcrafted, task-specific policies or are designed using full quantum-state information that is unavailable in real experiments and becomes impractical for large many-body systems.
Here we develop an adaptive measurement--feedback protocol based on reinforcement learning under partial observability. The controller uses only the history of experimentally accessible measurement outcomes to choose both the measurement operator and the feedback action in real time. To make training compatible with experiments, we introduce a stochastic terminal reward built from one-shot measurements of randomly sampled Hamiltonian components, avoiding unphysical full-state reconstruction while remaining an unbiased estimator of the target energy. We demonstrate the method by preparing ground states of the Bose--Hubbard model and by generating GHZ states, establishing a scalable and hardware-compatible route to quantum state preparation.
\end{abstract}

\maketitle

\textit{Introduction.}---Preparing many-body ground states is a central task across quantum science, underpinning quantum simulation of correlated phases, the extraction of ground-state properties, and the generation of structured entanglement. Beyond its role in condensed-matter and AMO settings, ground-state preparation also interfaces naturally with quantum information: many resource states and code manifolds can be realized as ground spaces of engineered Hamiltonians. Developing practical, scalable protocols for preparing such target ground states in many-body systems is therefore of broad interest.

Common approaches face obstacles in practice: adiabatic ramps slow down near small gaps and critical points~\cite{ProgrammableQuantumAnnealing@qiu.2020,QuantumAdiabaticOptimization@bombieri.2025,UniversalQuantumOptimization@ye.2023}; engineered dissipation~\cite{EmbeddingQuantumManyBody@wang.2024} requires finely tuned jump operators; and variational schemes such as VQE~\cite{MoGVQEMultiobjectiveGenetic@chivilikhin.2020,TETRISADAPTVQEAdaptiveAlgorithm@anastasiou.2024,ConstrainedVariationalQuantum@ryabinkin.2019,QuantumComputationElectronic@parrish.2019,AcceleratedVariationalQuantum@wang.2019} demand heavy measurement overhead and contend with nontrivial optimization landscapes. These limitations motivate strategies that exploit native measurement and control primitives directly.

Measurement--feedback control turns measurement backaction into a control resource, steering the system toward a target manifold through real-time closed-loop actions. Under weak monitoring, the controller has access only to a noisy measurement record, and the choice of observable governs both information gain and backaction. Current protocols often rely on fixed, handcrafted feedback laws~\cite{FieldLockedFock@zhou.2012,SimulatingNonlinearDynamics@munoz-arias.2020,FeedbackInducedQuantumPhase@ivanov.2020,ContinuousRealTimeTracking@shankar.2019,QuantumFeedbackTheory@zhang.2017,QuantumTheoryContinuous@wiseman.1994,AppearanceDisappearanceQuantum@sudhir.2017,NishimorisCatStable@zhu.2023,HierarchyTopologicalOrder@tantivasadakarn.2023,DeterministicSqueezedStates@cox.2016,SqueezingEntanglementDensity@wade.2015,UnconditionalQuantumNoiseSuppression@inoue.2013,DeterministicEntanglementSuperconducting@riste.2013,RealtimeQuantumFeedback@sayrin.2011,QuantumSuperpositionState@negretti.2007,QuantumFeedbackControl@yanagisawa.2006,BoundingFidelityQuantum@oconnor.2025,PreparingQuantumStates@wu.2023} that demand problem-specific intuition and degrade outside their design regime. Recent machine-learning approaches~\cite{ReinforcementLearningQuantum@bukov.2026,QuantumCircuitDiscovery@zen.2025a,QuantumOptimalControl@koch.2022} learn adaptive policies, but many train on signals available only in simulation, such as the full state or dense energy traces~\cite{ControllingNonergodicityQuantum@ye.2024,ManipulationSpinDynamics@chen.2019,TakingGradientsExperiments@august.2018,StrategyPreparingQuantum@zhao.2024,ClassifyingGlobalState@haug.2020,ReinforcementLearningQuantum@ernst.2025,FasterStatePreparation@guo.2021,TutorialOptimalControl@giannelli.2022,TamingQuantumSystems@duncan.2025,ReinforcementLearningOptimization@erdman.2024,ArtificiallyIntelligentMaxwells@erdman.2025,DeepReinforcementLearning@porotti.2022,MachineLearningGround@wang.2025}. Even terminal objectives are constrained: incompatible measurement settings across Hamiltonian terms mean a single shot can only reveal partial energy information.

In this work, we develop an adaptive measurement--feedback framework that is compatible with experimental observables at every stage of the learning loop \cite{ReinforcementLearningAutonomous@bukov.2018,RealtimeQuantumError@sivak.2023,ModelFreeQuantumControl@sivak.2022,RealizingDeepReinforcement@reuer.2023}. We cast ground-state preparation under weak monitoring as a partially observable control problem and train a history-dependent policy (implemented with a recurrent network) that acts only on the measurement record available in real time. Crucially, the policy jointly selects the measured collective observable and the feedback action on the fly, enabling it to balance information gain and measurement backaction across different dynamical regimes. To avoid simulation-only training signals, we design a terminal reward built from single-shot measurements of randomly sampled Hamiltonian terms, with an unbiased weighting that targets the total energy while respecting incompatible measurement settings. Applying this approach to interacting many-body models, we find that the learned closed-loop protocol robustly drives the system to substantially lower energies than fixed or handcrafted measurement--feedback baselines. Finally, by engineering target Hamiltonians whose ground spaces encode entangled resources, the same strategy provides a Hamiltonian-based route to preparing states such as GHZ states within the same measurement--feedback paradigm.

\begin{figure*}[t]
    \centering
    \includegraphics[width=0.85\textwidth]{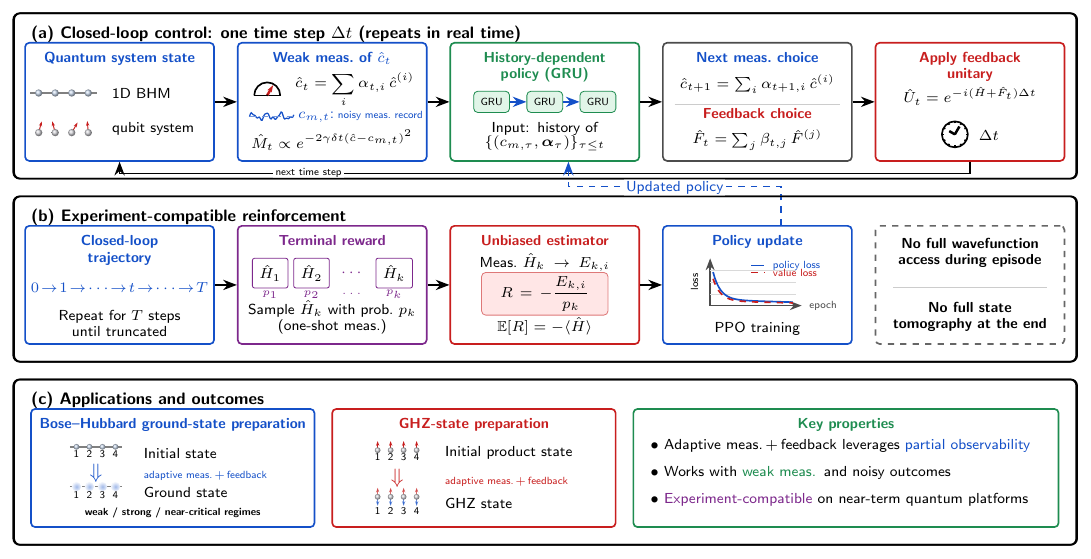}
    \caption{Schematic of the measurement--feedback reinforcement-learning framework. (a)~Closed-loop control: at each time step the system is weakly measured, a recurrent policy selects the next measurement and feedback action from the measurement record, and a feedback unitary is applied. (b)~Experiment-compatible reinforcement: a terminal reward is constructed from one-shot measurements of randomly sampled Hamiltonian terms, providing an unbiased energy estimator for PPO training. (c)~Applications: Bose--Hubbard ground-state preparation and GHZ-state generation.}
    \label{fig:sketch}
\end{figure*}

\textit{Measurement--feedback control process.}---We treat the measurement--feedback dynamics in a discretized form, partitioning time into short intervals of duration $\delta t$. During each interval, the system is weakly measured with respect to the observable $\hat{c}_t$, and a feedback unitary generated by $\hat{F}_t$ is applied. A weak measurement of the Hermitian observable $\hat{c}_t$ with measurement strength $\gamma$ is represented by the Kraus operator~\cite{StraightforwardIntroductionContinuous@jacobs.2006}:
\begin{equation}
\hat{M}_t(c_{m,t}) = \left(\frac{4\gamma \delta t}{\pi}\right)^{1/4} e^{-2\gamma \delta t (\hat{c}_t - c_{m,t})^2},
\end{equation}
where $c_{m,t}$ is the corresponding noisy measurement outcome that follows a normal distribution:
\begin{equation}
P(c_{m,t}) \sim \mathcal{N}\left(\mu=\langle \hat{c}_t \rangle,\sigma^2= \frac{1}{8 \gamma \delta t}\right).
\end{equation}
The parameter $\gamma$ controls the trade-off between information gain and measurement backaction: increasing $\gamma$ improves the precision of the measurement outcome but also enhances the disturbance to the quantum state.

Based on the measurement result $c_{m,t}$, we choose the feedback operator $\hat{F}_t$ according to our policy and apply it to modify the system's evolution. The whole time-evolution unitary operator is given by
\begin{equation}
\hat{U}_t = e^{-i(\hat{H} + \hat{F}_t) \delta t},
\end{equation}
where $\hat{H}$ is the original system Hamiltonian. The full update of the system state $|\psi(t)\rangle$ over one time step is then
\begin{equation}
|\psi(t + \delta t)\rangle \propto \hat{U}_t \hat{M}_t(c_{m,t}) |\psi(t)\rangle.
\end{equation}

\textit{Reinforcement learning.}---Our objective is to start from an experimentally accessible initial state---such as a product state or a fully polarized configuration---and drive the system toward its ground state through a measurement--feedback control process guided by a learned policy, as illustrated in Fig.~\ref{fig:sketch}. We cast this task as a partially observable Markov decision process (POMDP), in which the controller receives only the noisy stream of measurement outcomes rather than full knowledge of the quantum state. The measurement observable $\hat c_t$ and feedback operator $\hat F_t$ are parameterized in fixed operator bases,
\begin{equation}
\hat c_t=\sum_i \alpha_{t,i}\,\hat c^{(i)}, \qquad
\hat F_t=\sum_i \beta_{t,i}\,\hat F^{(i)},
\end{equation}
where $\{\hat c^{(i)}\}$, $\{\hat F^{(i)}\}$ are fixed basis operators, so that the weight vectors $\boldsymbol{\alpha}_t$ and $\boldsymbol{\beta}_t$ fully specify the measurement and feedback actions at step $t$. A GRU recurrent network observes the measurement weights $\boldsymbol{\alpha}_t$ just used and the measurement outcome $c_{m,t}$, then outputs the feedback weights $\boldsymbol{\beta}_t$ and the next measurement weights $\boldsymbol{\alpha}_{t+1}$. Since the feedback evolution is deterministic after $c_{m,t}$ is registered, $\boldsymbol{\alpha}_{t+1}$ can be produced in the same forward pass.

To maintain experimental compatibility, the reward must also be accessible. Ideally it would be the negative energy expectation $\langle -\hat H\rangle$ at the final state, but this is infeasible within a single trajectory: expectation values require averaging over multiple trajectories. Moreover, non-commuting Hamiltonian terms require incompatible measurement settings—in the Bose--Hubbard model, the hopping term $\hat{H}_{\rm kin}$ is measured via time-of-flight imaging while the interaction term $\hat{H}_{\rm int}$ uses in-situ imaging. We therefore construct the terminal reward from a single randomly sampled term: writing $\hat H=\sum_k \hat H_k$, one term $\hat H_k$ is chosen with probability $p_k$ and measured, yielding eigenvalue $E_{ki}$. The importance-weighted reward
\begin{equation}
R=-\frac{1}{p_k}E_{ki}
\end{equation}
is an unbiased estimator of the negative total energy, since $\mathbb{E}[R]=-\sum_k \langle \hat H_k\rangle=-\langle \hat H\rangle$.

To improve training stability, we reduce the variance of this stochastic reward in two steps. First, we center each term at its target ground-state expectation, defining $\tilde H_k=\hat H_k-\langle \hat H_k\rangle_0$ with $\langle\cdots\rangle_0$ the expectation in the target ground state; the reward $R=-(1/p_k)\tilde E_{ki}$ then has zero mean at the ground state for every sampled term, changing the objective only by a constant. Second, the variance $\mathrm{var}(R)=\sum_k \langle \tilde H_k^2\rangle_0/p_k$ is minimized under $\sum_k p_k=1$ by the choice $p_k\propto\sqrt{\langle \tilde H_k^2\rangle_0}$. Centering and optimal term sampling thus substantially suppress the variance while keeping the reward unbiased and experimentally compatible.

With this reward design, training of the measurement--feedback control policy is no longer confined to simulation environments that rely on privileged access to the full quantum state and are ultimately limited by the exponential growth of Hilbert space. Instead, the same training framework becomes compatible with experimental trajectories. We optimize the parameters of the recurrent policy using proximal policy optimization (PPO, implemented with the \texttt{PureJaxRL} library~\cite{DiscoveredPolicyOptimisation@lu.}), a stable policy-gradient method that limits excessively large updates between successive iterations. In each training round, the agent interacts with the measurement--feedback loop to collect trajectories, estimates the corresponding returns and advantages from the stochastic terminal reward, and updates the policy accordingly. Repeating this procedure yields a closed-loop measurement--feedback strategy that progressively drives the system toward the target low-energy state.

\textit{Numerical demonstrations under experimental constraints.}---We illustrate the proposed framework on two representative tasks: ground-state preparation in the Bose--Hubbard model (BHM) and GHZ-state preparation. The controller uses only the measurement outcomes available in experiment, demonstrating applicability to both many-body ground-state preparation and entangled-state generation.

\begin{figure}[!h]
    \centering
    \includegraphics[width=0.9\linewidth]{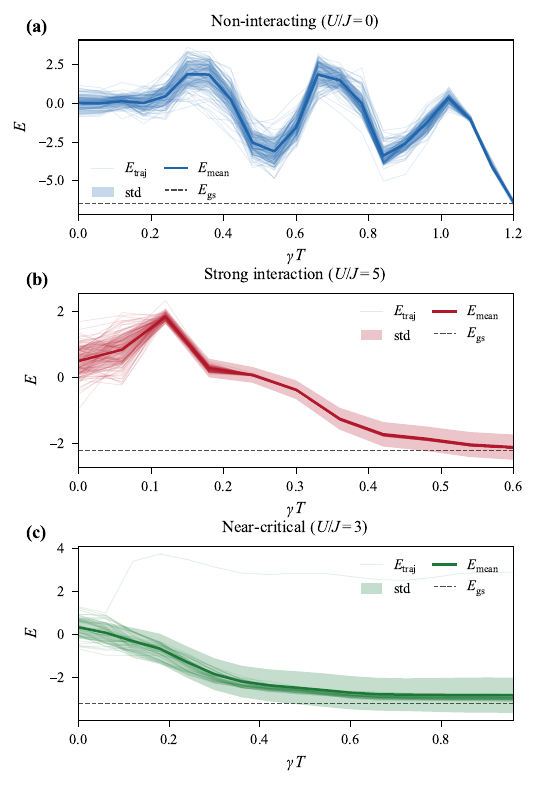}
    \caption{Energy evolution during measurement--feedback ground-state preparation of the four-site Bose--Hubbard model at unit filling, in the (a)~non-interacting ($U/J=0$), (b)~strong-interaction ($U/J=5$), and (c)~near-critical ($U/J=3$) regimes, with measurement strength $\gamma/J=0.3$. Thin colored lines show 100 sampled trajectories, the bold line is the ensemble mean, and the shaded region indicates $\pm 1$ standard deviation over 1000 episodes. The dashed line marks the exact ground-state energy $E_{\mathrm{gs}}$.}
    \label{fig:BHM_evolution}
\end{figure}

We first consider preparation of the ground state of the one-dimensional four-site Bose--Hubbard model at unit filling. The Hamiltonian contains only hopping and onsite interaction terms,
\begin{equation}
\hat H_{\mathrm{BHM}}
=
-J\sum_{i}\left(\hat a_i^\dagger \hat a_{i+1}+\mathrm{H.c.}\right)
+\frac{U}{2}\sum_i \hat n_i(\hat n_i-1).
\end{equation}
Following the operator choice of Wu \textit{et al.}~\cite{PreparingQuantumStates@wu.2023}, we take the measured observable and feedback operator to be
\begin{equation}
\hat c_t=\sum_{j}\alpha_{t,j}\,\hat n_j,
\quad
\hat F_t=(\beta_{t,1}+i\beta_{t,2})\sum_{j}\hat a_j^\dagger \hat a_{j+1}+\mathrm{H.c.},
\end{equation}
so that the policy adaptively chooses the density-weighted measurement profile and the complex hopping-feedback amplitude~\cite{MultipartiteEntangledSpatial@elliott.2015,DiffractionUnlimitedPositionMeasurement@ashida.2015,CavityAtomReview@Esslinger.2013}. We study three regimes—non-interacting, strong-interaction, and near-critical—at fixed measurement strength $\gamma/J=0.3$, initializing in the unit-filling product state $|1,1,1,1\rangle$ with a $10\%$ admixture of single particle--hole excitations to model preparation imperfections. As shown in Fig.~\ref{fig:BHM_evolution}, the non-interacting case converges by $\gamma T = 1.2$, whereas the protocol of Wu \textit{et al.}~\cite{PreparingQuantumStates@wu.2023} requires $\gamma T > 3$ in the same setting. In the strong-interaction regime, the adaptive policy continues to lower the energy efficiently and substantially outperforms fixed-feedback protocols, which fail to reach comparably low energies in the corresponding parameter range. Near criticality, where control is generally more challenging, the learned policy remains robust and still converges efficiently. Together, these results show that the proposed framework can adapt across qualitatively different many-body regimes without changing the underlying control architecture.

We next consider GHZ-state preparation for quantum information. We study two four-qubit examples, both initialized in the fully polarized spin-up product state. The first target is a four-qubit GHZ state, $|\mathrm{GHZ}_4\rangle=\frac{1}{\sqrt{2}}\bigl(|\uparrow\uparrow\uparrow\uparrow\rangle+|\downarrow\downarrow\downarrow\downarrow\rangle\bigr)$, characterized by the stabilizers $Z_1Z_2$, $Z_2Z_3$, $Z_3Z_4$, and $X_1X_2X_3X_4$, with parent Hamiltonian given by minus their sum, thereby testing the generation of genuine four-partite entanglement. The second target is a product of two two-qubit GHZ states on qubits $1,2$ and $3,4$, $|\mathrm{GHZ}_2\otimes\mathrm{GHZ}_2\rangle=\tfrac{1}{2}\bigl(|\uparrow\uparrow\rangle+|\downarrow\downarrow\rangle\bigr)_{12}\otimes\bigl(|\uparrow\uparrow\rangle+|\downarrow\downarrow\rangle\bigr)_{34}$, characterized by the stabilizers $Z_1Z_2$, $X_1X_2$, $Z_3Z_4$, and $X_3X_4$, again with parent Hamiltonian defined as minus their sum, thereby testing whether the same framework can simultaneously stabilize two independent entangled pairs.

Unlike the Bose--Hubbard case, where the physical Hamiltonian drives the evolution through $\hat H_{\rm BHM}+\hat F_t$, the GHZ parent Hamiltonian only defines the target and terminal reward; it is not applied during the closed-loop dynamics. The actual unitary feedback step is therefore $\hat U_t=e^{-i\hat F_t\delta t}$ with no native drift Hamiltonian.

\begin{figure}[!h]
    \centering
    \includegraphics[width=0.9\linewidth]{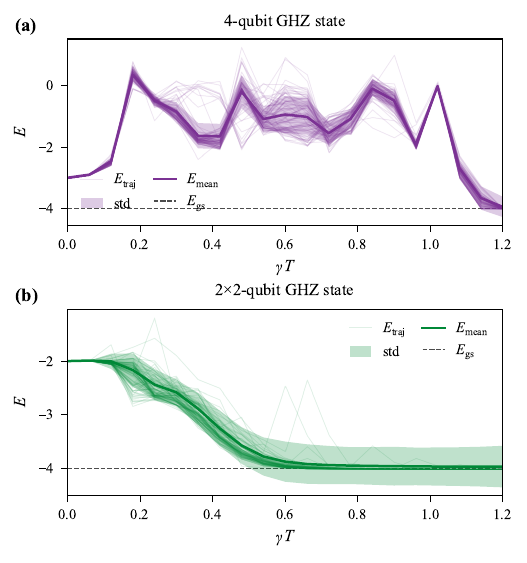}
    \caption{Energy evolution during GHZ-state preparation via measurement--feedback control, for (a)~the four-qubit GHZ state $|\mathrm{GHZ}_4\rangle$ and (b)~the product $|\mathrm{GHZ}_2\otimes\mathrm{GHZ}_2\rangle$ of two two-qubit GHZ states, with measurement strength $\gamma=0.3$. The same plotting conventions as Fig.~\ref{fig:BHM_evolution} are used. In both cases, the adaptive policy using only single-qubit $Z$-type measurements and $Y$-type feedback drives the energy close to $E_{\mathrm{gs}}=-4$, preparing the target entangled states with high fidelity.}
    \label{fig:GHZ_preparation}
\end{figure}

For these GHZ examples, we choose $\hat c_t=\sum_i \alpha_{t,i}\, Z_i$ and $\hat F_t=\sum_i \beta_{t,i}\, Y_i$,
so that the controller combines $Z$-type weak measurements with $Y$-type feedback rotations at fixed strength $\gamma=0.3$. Both $\hat c_t$ and $\hat F_t$ are sums of single-qubit operators, yet the protocol still prepares highly entangled states via collective measurement backaction. As shown in Fig.~\ref{fig:GHZ_preparation}, in both cases the energy reaches close to $E_{\mathrm{gs}}=-4$, indicating high-fidelity preparation. In this sense, the mechanism differs from conventional circuit-based preparation, where GHZ states are typically built using two-qubit entangling operations such as CNOT gates, which in platforms such as neutral-atom arrays rely on direct interactions, e.g., Rydberg blockade. The parent Hamiltonians separate into $Z$- and $X$-type components requiring different measurement bases, playing the role of the distinct terms $\hat H_k$ above. The same framework thus extends from many-body ground-state preparation to entangled-state generation for quantum computing.

\textit{Summary.}---We developed an adaptive measurement--feedback framework for quantum state preparation that operates entirely on experimentally accessible signals: a recurrent policy, trained as a POMDP, jointly selects the measurement observable and feedback action from the measurement record, and a stochastic terminal reward built from a single randomly sampled Hamiltonian term provides an unbiased energy estimator without full-state access. Numerical demonstrations show that the learned policy prepares Bose--Hubbard ground states across different interaction regimes, and stabilizes GHZ states as ground states of suitable parent Hamiltonians, establishing a scalable, hardware-compatible route to both many-body ground-state preparation and entangled resource-state generation, with natural extensions to broader closed-loop control tasks in quantum computing and error correction.

\textit{Acknowledgments.}---We thank Yadong Wu and Pengfei Zhang for helpful discussions.

\bibliographystyle{unsrt}
\bibliography{MFCPbib}

@article{AcceleratedVariationalQuantum@wang.2019,
  title = {Accelerated Variational Quantum Eigensolver},
  author = {Wang, Daochen and Higgott, Oscar and Brierley, Stephen},
  year = 2019,
  month = apr,
  journal = {Phys. Rev. Lett.},
  volume = {122},
  number = {14},
  pages = {140504},
  publisher = {American Physical Society},
  doi = {10.1103/PhysRevLett.122.140504},
  urldate = {2026-05-05}
}

@article{AppearanceDisappearanceQuantum@sudhir.2017,
  title = {Appearance and Disappearance of Quantum Correlations in Measurement-Based Feedback Control of a Mechanical Oscillator},
  author = {Sudhir, V. and Wilson, D. J. and Schilling, R. and Sch{\"u}tz, H. and Fedorov, S. A. and Ghadimi, A. H. and Nunnenkamp, A. and Kippenberg, T. J.},
  year = 2017,
  month = jan,
  journal = {Phys. Rev. X},
  volume = {7},
  number = {1},
  pages = {011001},
  publisher = {American Physical Society},
  doi = {10.1103/PhysRevX.7.011001},
  urldate = {2026-05-05}
}

@article{ArtificiallyIntelligentMaxwells@erdman.2025,
  title = {Artificially Intelligent Maxwell's Demon for Optimal Control of Open Quantum Systems},
  author = {Erdman, Paolo A and Czupryniak, Robert and Bhandari, Bibek and Jordan, Andrew N and No{\'e}, Frank and Eisert, Jens and Guarnieri, Giacomo},
  year = 2025,
  month = mar,
  journal = {Quantum Sci. Technol.},
  volume = {10},
  number = {2},
  pages = {025047},
  publisher = {IOP Publishing},
  doi = {10.1088/2058-9565/adbccf},
  urldate = {2026-05-05},
  langid = {english}
}

@article{BoundingFidelityQuantum@oconnor.2025,
  title = {Bounding Fidelity in Quantum Feedback Control: Theory and Applications to Dicke State Preparation},
  shorttitle = {Bounding Fidelity in Quantum Feedback Control},
  author = {O'Connor, Eoin and Ma, Hailan and Genoni, Marco G},
  year = 2025,
  month = jun,
  journal = {Quantum Sci. Technol.},
  volume = {10},
  number = {3},
  pages = {035049},
  publisher = {IOP Publishing},
  doi = {10.1088/2058-9565/ade55f},
  urldate = {2026-05-05},
  langid = {english}
}

@article{CavityAtomReview@Esslinger.2013,
  title = {Cold Atoms in Cavity-Generated Dynamical Optical Potentials},
  author = {Ritsch, Helmut and Domokos, Peter and Brennecke, Ferdinand and Esslinger, Tilman},
  year = 2013,
  month = apr,
  journal = {Rev. Mod. Phys.},
  volume = {85},
  number = {2},
  pages = {553--601},
  doi = {10.1103/RevModPhys.85.553},
  urldate = {2023-08-06},
  langid = {english}
}

@article{ClassifyingGlobalState@haug.2020,
  title = {Classifying Global State Preparation via Deep Reinforcement Learning},
  author = {Haug, Tobias and Mok, Wai-Keong and You, Jia-Bin and Zhang, Wenzu and Eng Png, Ching and Kwek, Leong-Chuan},
  year = 2020,
  month = dec,
  journal = {Mach. Learn.: Sci. Technol.},
  volume = {2},
  number = {1},
  pages = {01LT02},
  publisher = {IOP Publishing},
  doi = {10.1088/2632-2153/abc81f},
  urldate = {2026-05-05},
  langid = {english}
}

@article{ConstrainedVariationalQuantum@ryabinkin.2019,
  title = {Constrained Variational Quantum Eigensolver: Quantum Computer Search Engine in the Fock Space},
  shorttitle = {Constrained Variational Quantum Eigensolver},
  author = {Ryabinkin, Ilya G. and Genin, Scott N. and Izmaylov, Artur F.},
  year = 2019,
  month = jan,
  journal = {J. Chem. Theory Comput.},
  volume = {15},
  number = {1},
  pages = {249--255},
  publisher = {American Chemical Society},
  doi = {10.1021/acs.jctc.8b00943},
  urldate = {2026-05-05}
}

@article{ContinuousRealTimeTracking@shankar.2019,
  title = {Continuous Real-Time Tracking of a Quantum Phase Below the Standard Quantum Limit},
  author = {Shankar, Athreya and Greve, Graham P. and Wu, Baochen and Thompson, James K. and Holland, Murray},
  year = 2019,
  month = jun,
  journal = {Phys. Rev. Lett.},
  volume = {122},
  number = {23},
  pages = {233602},
  publisher = {American Physical Society},
  doi = {10.1103/PhysRevLett.122.233602},
  urldate = {2026-05-05}
}

@misc{ControllingNonergodicityQuantum@ye.2024,
  title = {Controlling Nonergodicity in Quantum Many-Body Systems by Reinforcement Learning},
  author = {Ye, Li-Li and Lai, Ying-Cheng},
  year = 2024,
  month = aug,
  journal = {arXiv.org},
  urldate = {2026-05-05},
  howpublished = {https://arxiv.org/abs/2408.11989v3},
  langid = {english}
}

@article{DeepReinforcementLearning@porotti.2022,
  title = {Deep Reinforcement Learning for Quantum State Preparation with Weak Nonlinear Measurements},
  author = {Porotti, Riccardo and Essig, Antoine and Huard, Benjamin and Marquardt, Florian},
  year = 2022,
  month = jun,
  journal = {Quantum},
  volume = {6},
  pages = {747},
  publisher = {Verein zur F\"orderung des Open Access Publizierens in den Quantenwissenschaften},
  doi = {10.22331/q-2022-06-28-747},
  urldate = {2026-05-05},
  langid = {british}
}

@article{DeterministicEntanglementSuperconducting@riste.2013,
  title = {Deterministic Entanglement of Superconducting Qubits by Parity Measurement and Feedback},
  author = {Rist{\`e}, D. and Dukalski, M. and Watson, C. A. and {de Lange}, G. and Tiggelman, M. J. and Blanter, Ya M. and Lehnert, K. W. and Schouten, R. N. and DiCarlo, L.},
  year = 2013,
  month = oct,
  journal = {Nature},
  volume = {502},
  number = {7471},
  pages = {350--354},
  publisher = {Nature Publishing Group},
  doi = {10.1038/nature12513},
  urldate = {2026-05-05},
  copyright = {2013 Springer Nature Limited},
  langid = {english}
}

@article{DeterministicSqueezedStates@cox.2016,
  title = {Deterministic Squeezed States with Collective Measurements and Feedback},
  author = {Cox, Kevin C. and Greve, Graham P. and Weiner, Joshua M. and Thompson, James K.},
  year = 2016,
  month = mar,
  journal = {Phys. Rev. Lett.},
  volume = {116},
  number = {9},
  pages = {093602},
  publisher = {American Physical Society},
  doi = {10.1103/PhysRevLett.116.093602},
  urldate = {2026-05-05}
}

@article{DiffractionUnlimitedPositionMeasurement@ashida.2015,
  title = {Diffraction-Unlimited Position Measurement of Ultracold Atoms in an Optical Lattice},
  author = {Ashida, Yuto and Ueda, Masahito},
  year = 2015,
  month = aug,
  journal = {Phys. Rev. Lett.},
  volume = {115},
  number = {9},
  pages = {095301},
  publisher = {American Physical Society},
  doi = {10.1103/PhysRevLett.115.095301},
  urldate = {2026-05-05}
}

@article{DiscoveredPolicyOptimisation@lu.,
  title = {Discovered Policy Optimisation},
  author = {Lu, Chris and Kuba, Jakub Grudzien and Letcher, Alistair and Metz, Luke and {de Witt}, Christian Schroeder and Foerster, Jakob},
  langid = {english}
}

@article{EmbeddingQuantumManyBody@wang.2024,
  title = {Embedding Quantum Many-Body Scars into Decoherence-Free Subspaces},
  author = {Wang, He-Ran and Yuan, Dong and Zhang, Shun-Yao and Wang, Zhong and Deng, Dong-Ling and Duan, L.-M.},
  year = 2024,
  month = apr,
  journal = {Phys. Rev. Lett.},
  volume = {132},
  number = {15},
  pages = {150401},
  doi = {10.1103/PhysRevLett.132.150401},
  urldate = {2026-05-05},
  langid = {english}
}

@article{FasterStatePreparation@guo.2021,
  title = {Faster State Preparation across Quantum Phase Transition Assisted by Reinforcement Learning},
  author = {Guo, Shuai-Feng and Chen, Feng and Liu, Qi and Xue, Ming and Chen, Jun-Jie and Cao, Jia-Hao and Mao, Tian-Wei and Tey, Meng Khoon and You, Li},
  year = 2021,
  month = feb,
  journal = {Phys. Rev. Lett.},
  volume = {126},
  number = {6},
  pages = {060401},
  publisher = {American Physical Society},
  doi = {10.1103/PhysRevLett.126.060401},
  urldate = {2026-05-05}
}

@article{FeedbackInducedQuantumPhase@ivanov.2020,
  title = {Feedback-Induced Quantum Phase Transitions Using Weak Measurements},
  author = {Ivanov, D. A. and Ivanova, T. {\relax Yu}. and {Caballero-Benitez}, S. F. and Mekhov, I. B.},
  year = 2020,
  month = jan,
  journal = {Phys. Rev. Lett.},
  volume = {124},
  number = {1},
  pages = {010603},
  publisher = {American Physical Society},
  doi = {10.1103/PhysRevLett.124.010603},
  urldate = {2026-05-05}
}

@article{FieldLockedFock@zhou.2012,
  title = {Field Locked to a Fock State by Quantum Feedback with Single Photon Corrections},
  author = {Zhou, X. and Dotsenko, I. and Peaudecerf, B. and Rybarczyk, T. and Sayrin, C. and Gleyzes, S. and Raimond, J. M. and Brune, M. and Haroche, S.},
  year = 2012,
  month = jun,
  journal = {Phys. Rev. Lett.},
  volume = {108},
  number = {24},
  pages = {243602},
  publisher = {American Physical Society},
  doi = {10.1103/PhysRevLett.108.243602}
}

@article{HierarchyTopologicalOrder@tantivasadakarn.2023,
  title = {Hierarchy of Topological Order From Finite-Depth Unitaries, Measurement, and Feedforward},
  author = {Tantivasadakarn, Nathanan and Vishwanath, Ashvin and Verresen, Ruben},
  year = 2023,
  month = jun,
  journal = {PRX Quantum},
  volume = {4},
  number = {2},
  pages = {020339},
  publisher = {American Physical Society},
  doi = {10.1103/PRXQuantum.4.020339},
  urldate = {2026-05-05}
}

@misc{MachineLearningGround@wang.2025,
  title = {Machine Learning for Ground State Preparation via Measurement and Feedback},
  author = {Wang, Chuanxin and You, Yi-Zhuang},
  year = 2025,
  month = feb,
  number = {arXiv:2502.06517},
  eprint = {2502.06517},
  primaryclass = {quant-ph},
  publisher = {arXiv},
  doi = {10.48550/arXiv.2502.06517},
  urldate = {2026-05-05},
  archiveprefix = {arXiv}
}

@misc{ManipulationSpinDynamics@chen.2019,
  title = {Manipulation of Spin Dynamics by Deep Reinforcement Learning Agent},
  author = {Chen, Jun-Jie and Xue, Ming},
  year = 2019,
  month = jan,
  journal = {arXiv.org},
  urldate = {2026-05-05},
  howpublished = {https://arxiv.org/abs/1901.08748v2},
  langid = {english}
}

@article{ModelFreeQuantumControl@sivak.2022,
  title = {Model-Free Quantum Control with Reinforcement Learning},
  author = {Sivak, V. V. and Eickbusch, A. and Liu, H. and Royer, B. and Tsioutsios, I. and Devoret, M. H.},
  year = 2022,
  month = mar,
  journal = {Phys. Rev. X},
  volume = {12},
  number = {1},
  pages = {011059},
  publisher = {American Physical Society},
  doi = {10.1103/PhysRevX.12.011059},
  urldate = {2026-05-05}
}

@article{MoGVQEMultiobjectiveGenetic@chivilikhin.2020,
  title = {MoG-VQE: Multiobjective Genetic Variational Quantum Eigensolver},
  shorttitle = {MoG-VQE},
  author = {Chivilikhin, D. and Samarin, A. and Ulyantsev, V. and Iorsh, I. and Oganov, A. and Kyriienko, O.},
  year = 2020,
  month = jul,
  journal = {arXiv: Quantum Physics},
  urldate = {2026-05-05}
}

@article{MultipartiteEntangledSpatial@elliott.2015,
  title = {Multipartite Entangled Spatial Modes of Ultracold Atoms Generated and Controlled by Quantum Measurement},
  author = {Elliott, T. J. and Kozlowski, W. and {Caballero-Benitez}, S. F. and Mekhov, I. B.},
  year = 2015,
  month = mar,
  journal = {Phys. Rev. Lett.},
  volume = {114},
  number = {11},
  pages = {113604},
  publisher = {American Physical Society},
  doi = {10.1103/PhysRevLett.114.113604},
  urldate = {2026-05-05}
}

@article{NishimorisCatStable@zhu.2023,
  title = {Nishimori's Cat: Stable Long-Range Entanglement from Finite-Depth Unitaries and Weak Measurements},
  shorttitle = {Nishimori's Cat},
  author = {Zhu, Guo-Yi and Tantivasadakarn, Nathanan and Vishwanath, Ashvin and Trebst, Simon and Verresen, Ruben},
  year = 2023,
  month = nov,
  journal = {Phys. Rev. Lett.},
  volume = {131},
  number = {20},
  pages = {200201},
  publisher = {American Physical Society},
  doi = {10.1103/PhysRevLett.131.200201},
  urldate = {2026-05-05}
}

@article{PreparingQuantumStates@wu.2023,
  title = {Preparing Quantum States by Measurement-Feedback Control with Bayesian Optimization},
  author = {Wu, Yadong and Yao, Juan and Zhang, Pengfei},
  year = 2023,
  month = jul,
  journal = {Front. Phys.},
  volume = {18},
  number = {6},
  pages = {61301},
  doi = {10.1007/s11467-023-1311-5},
  urldate = {2026-05-05},
  langid = {english}
}

@article{ProgrammableQuantumAnnealing@qiu.2020,
  title = {Programmable Quantum Annealing Architectures with Ising Quantum Wires},
  author = {Qiu, Xingze and Zoller, Peter and Li, Xiaopeng},
  year = 2020,
  month = nov,
  journal = {PRX Quantum},
  volume = {1},
  number = {2},
  pages = {020311},
  doi = {10.1103/PRXQuantum.1.020311},
  urldate = {2024-12-17},
  langid = {english}
}

@article{QuantumAdiabaticOptimization@bombieri.2025,
  title = {Quantum Adiabatic Optimization with Rydberg Arrays: Localization Phenomena and Encoding Strategies},
  shorttitle = {Quantum Adiabatic Optimization with Rydberg Arrays},
  author = {Bombieri, Lisa and Zeng, Zhongda and Tricarico, Roberto and Lin, Rui and Notarnicola, Simone and Cain, Madelyn and Lukin, Mikhail D. and Pichler, Hannes},
  year = 2025,
  month = apr,
  journal = {PRX Quantum},
  volume = {6},
  number = {2},
  pages = {020306},
  publisher = {American Physical Society},
  doi = {10.1103/PRXQuantum.6.020306},
  urldate = {2026-05-05}
}

@article{QuantumCircuitDiscovery@zen.2025a,
  title = {Quantum Circuit Discovery for Fault-Tolerant Logical State Preparation with Reinforcement Learning},
  author = {Zen, Remmy and Olle, Jan and Colmenarez, Luis and Puviani, Matteo and M{\"u}ller, Markus and Marquardt, Florian},
  year = 2025,
  month = oct,
  journal = {Phys. Rev. X},
  volume = {15},
  number = {4},
  pages = {041012},
  doi = {10.1103/gqpr-dgz7},
  urldate = {2026-06-11},
  langid = {english}
}

@article{QuantumComputationElectronic@parrish.2019,
  title = {Quantum Computation of Electronic Transitions Using a Variational Quantum Eigensolver},
  author = {Parrish, Robert M. and Hohenstein, Edward G. and McMahon, Peter L. and Mart{\'i}nez, Todd J.},
  year = 2019,
  month = jun,
  journal = {Phys. Rev. Lett.},
  volume = {122},
  number = {23},
  pages = {230401},
  publisher = {American Physical Society},
  doi = {10.1103/PhysRevLett.122.230401},
  urldate = {2026-05-05}
}

@article{QuantumFeedbackControl@yanagisawa.2006,
  title = {Quantum Feedback Control for Deterministic Entangled Photon Generation},
  author = {Yanagisawa, Masahiro},
  year = 2006,
  month = nov,
  journal = {Phys. Rev. Lett.},
  volume = {97},
  number = {19},
  pages = {190201},
  publisher = {American Physical Society},
  doi = {10.1103/PhysRevLett.97.190201},
  urldate = {2026-05-05}
}

@article{QuantumFeedbackTheory@zhang.2017,
  title = {Quantum Feedback: Theory, Experiments, and Applications},
  shorttitle = {Quantum Feedback},
  author = {Zhang, Jing and Liu, Yu-xi and Wu, Re-Bing and Jacobs, Kurt and Nori, Franco},
  year = 2017,
  month = mar,
  journal = {Physics Reports},
  series = {Quantum Feedback: Theory, Experiments, and Applications},
  volume = {679},
  pages = {1--60},
  doi = {10.1016/j.physrep.2017.02.003},
  urldate = {2026-05-05}
}

@article{QuantumOptimalControl@koch.2022,
  title = {Quantum Optimal Control in Quantum Technologies. Strategic Report on Current Status, Visions and Goals for Research in Europe},
  author = {Koch, Christiane P. and Boscain, Ugo and Calarco, Tommaso and Dirr, Gunther and Filipp, Stefan and Glaser, Steffen J. and Kosloff, Ronnie and Montangero, Simone and {Schulte-Herbr{\"u}ggen}, Thomas and Sugny, Dominique and Wilhelm, Frank K.},
  year = 2022,
  month = jul,
  journal = {EPJ Quantum Technol.},
  volume = {9},
  number = {1},
  pages = {19},
  doi = {10.1140/epjqt/s40507-022-00138-x},
  urldate = {2026-05-05},
  langid = {english}
}

@article{QuantumSuperpositionState@negretti.2007,
  title = {Quantum Superposition State Production by Continuous Observations and Feedback},
  author = {Negretti, Antonio and Poulsen, Uffe V. and M{\o}lmer, Klaus},
  year = 2007,
  month = nov,
  journal = {Phys. Rev. Lett.},
  volume = {99},
  number = {22},
  pages = {223601},
  publisher = {American Physical Society},
  doi = {10.1103/PhysRevLett.99.223601},
  urldate = {2026-05-05}
}

@article{QuantumTheoryContinuous@wiseman.1994,
  title = {Quantum Theory of Continuous Feedback},
  author = {Wiseman, H. M.},
  year = 1994,
  month = mar,
  journal = {Phys. Rev. A},
  volume = {49},
  number = {3},
  pages = {2133--2150},
  publisher = {American Physical Society},
  doi = {10.1103/PhysRevA.49.2133},
  urldate = {2026-05-05}
}

@article{RealizingDeepReinforcement@reuer.2023,
  title = {Realizing a Deep Reinforcement Learning Agent for Real-Time Quantum Feedback},
  author = {Reuer, Kevin and Landgraf, Jonas and F{\"o}sel, Thomas and O'Sullivan, James and Beltr{\'a}n, Liberto and Akin, Abdulkadir and Norris, Graham J. and Remm, Ants and Kerschbaum, Michael and Besse, Jean-Claude and Marquardt, Florian and Wallraff, Andreas and Eichler, Christopher},
  year = 2023,
  month = nov,
  journal = {Nat Commun},
  volume = {14},
  number = {1},
  pages = {7138},
  publisher = {Nature Publishing Group},
  doi = {10.1038/s41467-023-42901-3},
  urldate = {2026-05-05},
  copyright = {2023 The Author(s)},
  langid = {english}
}

@article{RealtimeQuantumError@sivak.2023,
  title = {Real-Time Quantum Error Correction beyond Break-Even},
  author = {Sivak, V. V. and Eickbusch, A. and Royer, B. and Singh, S. and Tsioutsios, I. and Ganjam, S. and Miano, A. and Brock, B. L. and Ding, A. Z. and Frunzio, L. and Girvin, S. M. and Schoelkopf, R. J. and Devoret, M. H.},
  year = 2023,
  month = apr,
  journal = {Nature},
  volume = {616},
  number = {7955},
  pages = {50--55},
  publisher = {Nature Publishing Group},
  doi = {10.1038/s41586-023-05782-6},
  urldate = {2026-05-05},
  copyright = {2023 The Author(s), under exclusive licence to Springer Nature Limited},
  langid = {english}
}

@article{RealtimeQuantumFeedback@sayrin.2011,
  title = {Real-Time Quantum Feedback Prepares and Stabilizes Photon Number States},
  author = {Sayrin, Cl{\'e}ment and Dotsenko, Igor and Zhou, Xingxing and Peaudecerf, Bruno and Rybarczyk, Th{\'e}o and Gleyzes, S{\'e}bastien and Rouchon, Pierre and Mirrahimi, Mazyar and Amini, Hadis and Brune, Michel and Raimond, Jean-Michel and Haroche, Serge},
  year = 2011,
  month = sep,
  journal = {Nature},
  volume = {477},
  number = {7362},
  pages = {73--77},
  publisher = {Nature Publishing Group},
  doi = {10.1038/nature10376},
  urldate = {2026-05-05},
  copyright = {2011 Springer Nature Limited},
  langid = {english}
}

@article{ReinforcementLearningAutonomous@bukov.2018,
  title = {Reinforcement Learning for Autonomous Preparation of Floquet-Engineered States: Inverting the Quantum Kapitza Oscillator},
  shorttitle = {Reinforcement Learning for Autonomous Preparation of Floquet-Engineered States},
  author = {Bukov, Marin},
  year = 2018,
  month = dec,
  journal = {Phys. Rev. B},
  volume = {98},
  number = {22},
  pages = {224305},
  publisher = {American Physical Society},
  doi = {10.1103/PhysRevB.98.224305},
  urldate = {2026-05-05}
}

@article{ReinforcementLearningOptimization@erdman.2024,
  title = {Reinforcement Learning Optimization of the Charging of a Dicke Quantum Battery},
  author = {Erdman, Paolo Andrea and Andolina, Gian Marcello and Giovannetti, Vittorio and No{\'e}, Frank},
  year = 2024,
  month = dec,
  journal = {Phys. Rev. Lett.},
  volume = {133},
  number = {24},
  pages = {243602},
  publisher = {American Physical Society},
  doi = {10.1103/PhysRevLett.133.243602},
  urldate = {2026-05-05}
}

@misc{ReinforcementLearningQuantum@bukov.2026,
  title = {Reinforcement Learning for Quantum Technology},
  author = {Bukov, Marin and Marquardt, Florian},
  year = 2026,
  month = jan,
  number = {arXiv:2601.18953},
  eprint = {2601.18953},
  primaryclass = {quant-ph},
  publisher = {arXiv},
  doi = {10.48550/arXiv.2601.18953},
  urldate = {2026-05-05},
  archiveprefix = {arXiv}
}

@misc{ReinforcementLearningQuantum@ernst.2025,
  title = {Reinforcement Learning for Quantum Control under Physical Constraints},
  author = {Ernst, Jan Ole and Chatterjee, Aniket and Franzmeyer, Tim and Kuhn, Axel},
  year = 2025,
  month = jan,
  journal = {arXiv.org},
  urldate = {2026-05-05},
  howpublished = {https://arxiv.org/abs/2501.14372v2},
  langid = {english}
}

@article{SimulatingNonlinearDynamics@munoz-arias.2020,
  title = {Simulating Nonlinear Dynamics of Collective Spins via Quantum Measurement and Feedback},
  author = {{Mu{\~n}oz-Arias}, Manuel H. and Poggi, Pablo M. and Jessen, Poul S. and Deutsch, Ivan H.},
  year = 2020,
  month = mar,
  journal = {Phys. Rev. Lett.},
  volume = {124},
  number = {11},
  pages = {110503},
  publisher = {American Physical Society},
  doi = {10.1103/PhysRevLett.124.110503},
  urldate = {2026-05-05}
}

@article{SqueezingEntanglementDensity@wade.2015,
  title = {Squeezing and Entanglement of Density Oscillations in a Bose-Einstein Condensate},
  author = {Wade, Andrew C. J. and Sherson, Jacob F. and M{\o}lmer, Klaus},
  year = 2015,
  month = aug,
  journal = {Phys. Rev. Lett.},
  volume = {115},
  number = {6},
  pages = {060401},
  publisher = {American Physical Society},
  doi = {10.1103/PhysRevLett.115.060401},
  urldate = {2026-05-05}
}

\end{document}